\documentclass{epl}

\usepackage[english]{babel}
\usepackage{amssymb}
\usepackage{amsmath}
\usepackage{amsfonts}

\title{Closed-loop control strategy with improved current for a flashing ratchet}
\shorttitle{Closed-loop control\ldots}
\author{Luis Dinis\inst{1}, Juan M.R.~Parrondo\inst{1} \and Francisco J.~Cao\inst{1,2}}
\institute{
 \inst{1} Grupo Interdisciplinar de Sistemas Complejos (GISC) and
Dept.~de F\'{\i}sica At\'{o}mica, Molecular y Nuclear,
 Universidad Complutense de Madrid, 28040-Madrid, Spain.\\
 \inst{2} LERMA, Observatoire de Paris, Laboratoire Associ\'e au
 CNRS UMR 8112, 61, Ave\-nue de l'Observatoire, 75015 Paris, France.
 }

\pacs{05.40.-a}{Fluctuation phenomena, random processes, noise,
and Brownian motion}
\pacs{02.30.Yy}{Control theory}

\begin{document}
 \maketitle
\begin{abstract}

We show how to switch on and off the ratchet potential of a collective Brownian motor, depending only on the position of the particles, in order to attain a current higher than or at least equal to that induced by any periodic flashing. Maximization of instant velocity turns out to be the optimal protocol for one particle but is nevertheless defeated by a periodic switching when a sufficiently large ensemble of particles is considered. The protocol presented in this letter, although not the optimal one, yields approximately the same current as the optimal protocol for one particle and as the optimal periodic switching for an infinite number of them.
\end{abstract}

\section{Introduction}
Brownian motors, acting as thermal fluctuation rectifiers, have been attracting considerable attention in the past ten years mainly due to their potential applications in biology, condensed matter and nanotechnology but also to their theoretical relevancy in statistical mechanics \cite{rei,lin}.
In most of the cases, rectification is achieved by means of an external periodic or random perturbation applied to an equilibrium system with some underlying spatial or temporal asymmetry.

However, it is only recently that research has been focused on introducing control in Brownian ratchets.
Tarlie and Astumian studied in \cite{tarast} the effect of a controllable time modulation of the applied sawtooth potential and find the  modulation that maximizes particle flow.
Other ways to control particle current have been proposed that are specially useful in systems where the ratchet potential parameters are not easily tunable, as in  most two dimensional ratchet setups where the Brownian particles interact with an asymmetric substrate. Examples of these devices are a rectifier of magnetic flux quanta in superconductors \cite{vil} or synthetic ion channels \cite{siw}.
According to Savel'ev {\em et al.}, flux in some of these ratchets can be controlled by addition of other species of particles that interact with the original ones \cite{savelevprl}. Alternatively, the application of two mixing signals could control ratchet-like particle transport of ferrofluids or dislocations in crystalline solids, pumping of electrons in quantum dots or colloids in arrays of optical tweezers \cite{savelevepl}.

All the control strategies mentioned so far share the property of being open-loop control policies, that is, they do not use any information about the state of the system. In contrast, we considered in ref.~\cite{cao} the idea of a feedback controlled perturbation, i.e., a modulation of the ratchet potential depending on the state of the system. Such a feedback control could be implemented in systems where particles are monitored, as in some experimental setups of colloidal particles \cite{rou,mar}.
The system described in ref.~\cite{cao} consisted on an ensemble of Brownian particles in a flashing ratchet potential that can be switched on or off depending on  the position of particles, with the aim of maximizing the current. We studied in detail a protocol that maximizes the instant velocity of the center of mass of the ensemble. That (closed-loop) protocol is optimal for one particle and performs better than any periodic (open-loop) flashing for ensembles of moderate sizes, but is defeated by random or periodic switching for large ensembles.
Furthermore, it can be shown that particle current vanishes as $N\to\infty$, even though the state of the potential is chosen so that the center of mass velocity is maximal at every instant.

In this letter we propose a modification of the aforementioned feedback control protocol that may switch off the potential even if that implies a loss of instantaneous velocity, or switch it on when that makes the center of mass initially recede. Nevertheless, this new protocol beats the original one for a sufficiently large number of particles. Although it might be surprising at first sight, this means you must stop or even move backwards at certain periods of time in order to move faster on average.

In contrast, maximization of the instant velocity protocol is a ``greedy'' strategy and will therefore never switch the potential off if particles are still moving forward or switch it on if this means an initial backward movement. In some sense, the new protocol performs  better by being less greedy, an effect also analyzed in \cite{cleu1,cleu2} for the case of paradoxical games,  which are closely related to the flashing ratchet.

\section{The model}
\begin{figure}
\begin{center}
\includegraphics[width=0.5\textwidth]{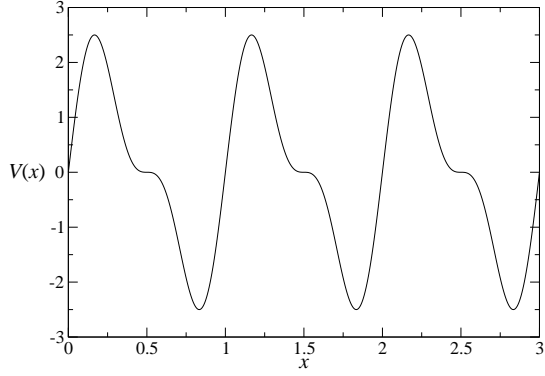}
\end{center}
\caption{\label{fig.pot} Ratchet potential $V(x)=\frac{2V_0}{3\sqrt{3}}\left(\sin(2\pi x/L)+\frac{1}{2}\sin(4\pi x/L)\right)$, for $V_0=5kT$ (Units: $kT=1$ and $L=1$).  }
\end{figure}

The model we will consider here consists of an ensemble of $N$ non-interacting overdamped Brownian particles at temperature $T$ in  a asymmetric periodic potential    that can be switched on or off. The dynamics is described by the Langevin equation
\begin{equation}
\gamma \dot x_i(t) = \alpha(t) F(x_i(t)) + \xi_i(t) \;; \quad\quad
i = 1 \ldots N, \label{lang1}
\end{equation}
 where $ x_i (t)$ the position of particle $ i $,
$\gamma$ is the friction  and  $ \xi_i (t)$ are thermal noises
with zero mean and correlation $ \langle \xi_i(t)\xi_j(t') \rangle
= 2 \gamma k T \delta_{ij} \delta(t-t')$. The force is given by
$F(x) = -V'(x)$. As the ratchet potential $V(x)$,  we will consider from now on the one depicted in fig.~\ref{fig.pot}, given by $V(x)=2V_0\left(\sin(2\pi x/L)+\frac{1}{2}\sin(4\pi x/L)\right)/3\sqrt{3}$, with $V_0=5kT$.
 We have found analogous results for other heights of the potential
 and for other smooth potential shapes.

Finally, $\alpha(t)$ is a control parameter which we assume that can take
on the values 1 and 0, i.e., the only allowed operations on the
Brownian motor consist in switching on and off the potential
$V(x)$.

\section{Control strategies}
The following two control strategies were previously compared in \cite{cao}, focusing on the induced current of particles.
\begin{itemize}

\item {\em Optimal periodic switching}: $ \alpha(t+\tau_{\text{on}}+\tau_{\text{off}})= \alpha(t)
$, with $ \alpha(t)=1 $ for $ t \in [0,\tau_{\text{on}})$, and $ \alpha(t)=0
$ for $ t \in [\tau_{\text{on}},\tau_{\text{on}}+\tau_{\text{off}})$. This case is equivalent to the
periodic flashing ratchet, since particles
are independent, although with different periods for the on and off steps.
In the following, the switching times $\tau_{\text{on}}$ and $\tau_{\text{off}}$ are set to achieve the highest stationary current. For $kT=1$ and $V_0=5kT$, these values are $\tau_{\text{on}}=0.06L^2/D$ and $\tau_{\text{off}}=0.05L^2/D$.
\item \emph{Maximization of center of mass instant velocity}:

\begin{equation} \label{alphacont}
\alpha(t) = \Theta (f(t))
\end{equation}
where
\begin{equation}
f(t)=\frac{1}{N}\sum_i^NF(x_i)\label{netforce}
\end{equation}
is the net force per particle if the ratchet potential were on, and $\Theta(y)$ is the
Heaviside function, $\Theta(y)=1$ if $y\geq 0$ and 0 otherwise.
As can be deduced from Eq.~\eqref{lang1}, this strategy maximizes the
instant velocity of the center of mass, $ \dot x_{cm}(t)=
\frac{1}{N} \sum_{i=1}^N \dot x_i(t)$.
\end{itemize}

\begin{figure}
\twoimages[width=0.45\textwidth]{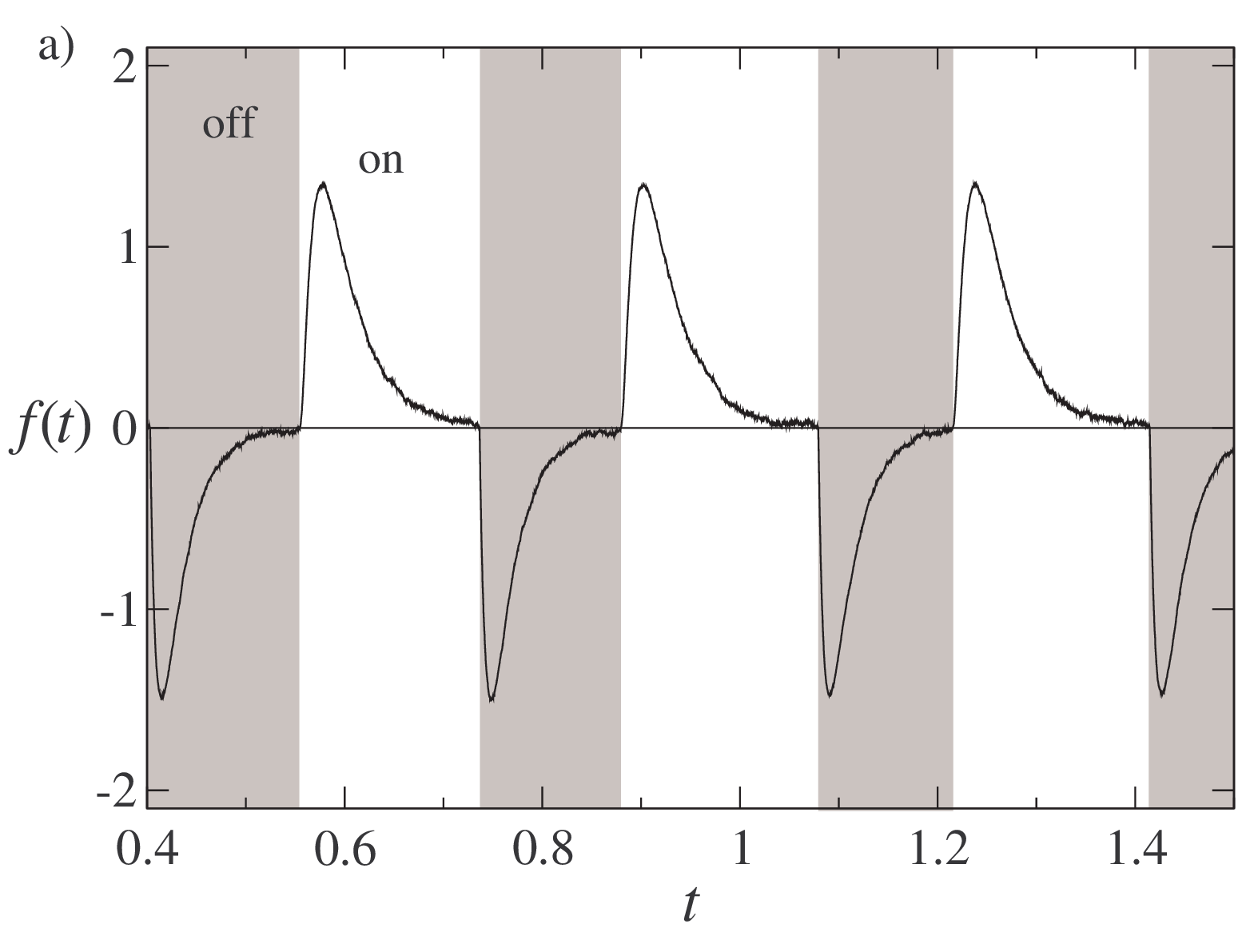}{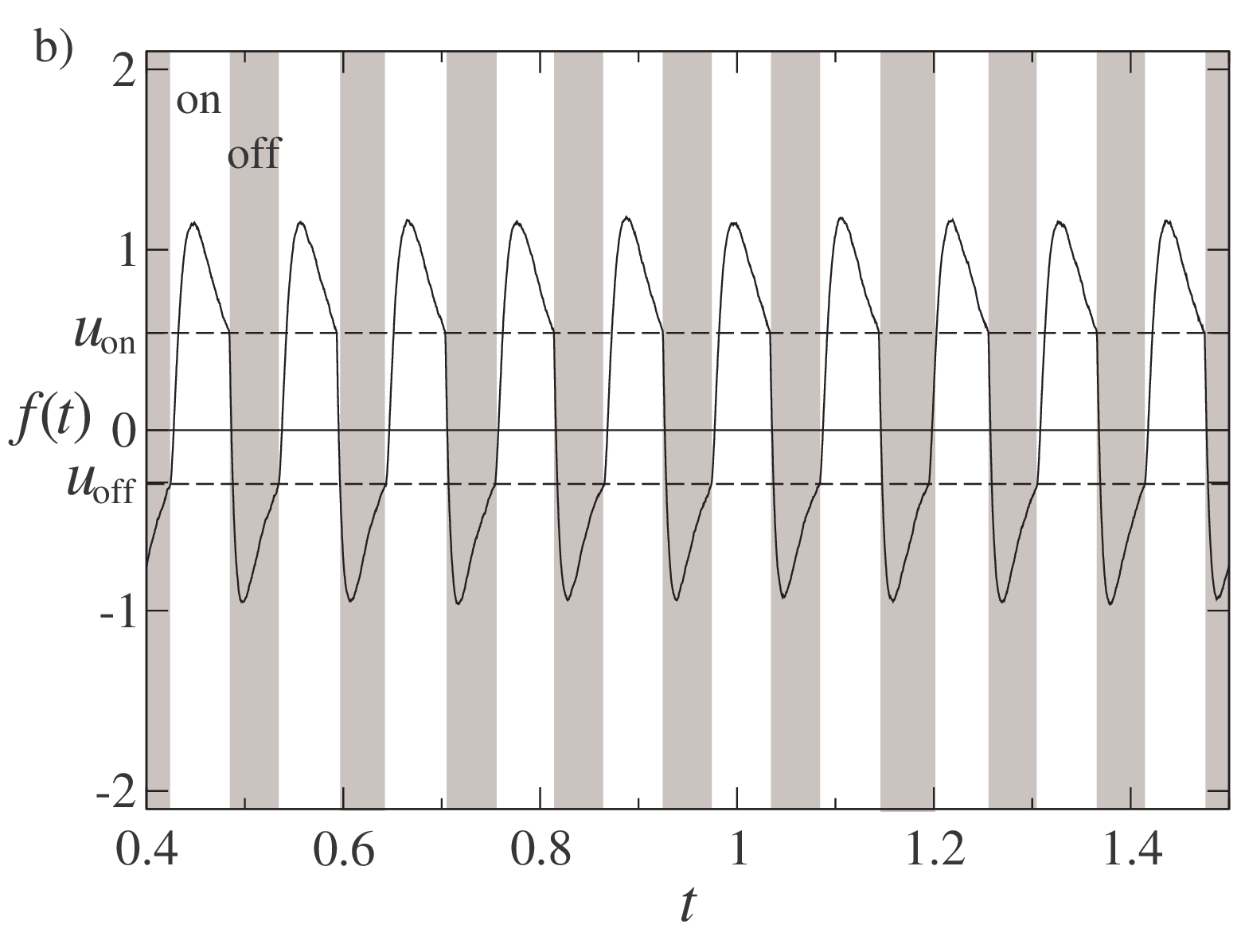}
\caption{\label{evolucion} Time dependence of the net force per particle $f(t)$ in simulations with $N=10^6$ for instant velocity of the center of mass maximization (a) and feedback control with thresholds (b). Thresholds are depicted as  dashed horizontal lines. }
\end{figure}

\begin{figure}
\begin{center}
\includegraphics[width=0.65\textwidth]{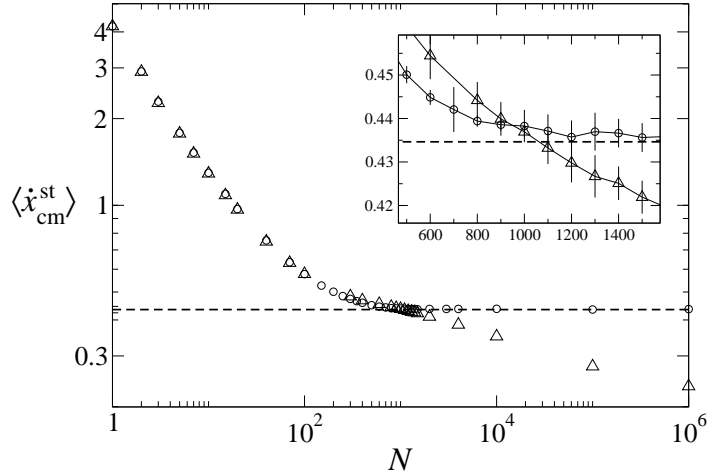}
\end{center}
\caption{\label{results} Simulation results of the average of the center of mass velocity $\langle\dot x_{\text{c.m.}}\rangle$, for the periodic switching with optimal periods $\tau_{\text{on}}=0.06L^2/D$ and $\tau_{\text{off}}=0.05L^2/D$ (dashed line); feedback control with thresholds (circles); and maximization of center of mass velocity (triangles). The inset shows the region where the strategies cross in more detail. A line through the points has been drawn to guide the eye. }
\end{figure}

As stated above this protocol is  the optimal one for one particle, performs better than any periodic switching only up to approximately a thousand particles (for our choice of the potential), and gives a vanishing current for $N\to\infty$ (see below fig.~\ref{results}).
For large $N$, any switching of the potential is induced by fluctuations in the force per particle around zero, as shown in fig.~\ref{evolucion}a. Since fluctuations vanish as $1/\sqrt{N}$, then  the system gets trapped in either the on or off state longer and longer times as $N$ increases, slowing down the switching that makes the ratchet work (see ref.~\cite{cao} for details).

This can be better understood in the limiting  case $N=\infty$, where the system evolves without fluctuations. The distribution of particles $\rho(x,t)$ coincides with the solution of the following mean-field Fokker-Planck equation:

\begin{equation}
\gamma \partial_t \rho (x,t) = \left[- \alpha(t)
\partial_x F(x) +kT\partial^2_{x}\right] \rho (x,t). \label{FP}
\end{equation}
Following \cite{cao}, evolution of the net force per particle $f(t)$ for sufficiently large $N$ would have two contributions: the evolution of $f^\infty(t)\equiv \langle F(x)\rangle$ where the average $\langle \cdot\rangle$ is taken over the solution $\rho(x,t)$ of Fokker-Planck eq.~\eqref{FP}, plus fluctuations of order $\sqrt{\langle F^2\rangle/N}$.
As explained above, the switching frequency is so slow that the system almost reaches the equilibrium distribution of either the on or off potential before a switch occurs. Hence, we can approximately compute $f^\infty(t)$ in two parts, one corresponding to either state of the potential. The ``on'' part $f^\infty_{\text{on}}(t)$ is computed by numerical integration of eq.~\eqref{FP} with initial condition given by the equilibrium distribution of the off potential and conversely, $f^\infty_{\text{off}}(t)$ with initial condition given by the equilibrium distribution of the on potential. These solutions $f^\infty_{\text{on}}(t)$, $f^\infty_{\text{off}}(t)$
would respectively represent, after adding the fluctuations, the cusps and valleys in fig.~\ref{evolucion}a.

Figure~\ref{evolucion} shows an initial increase or decrease of the force per particle right after every switch, followed by a tail that slowly decays towards zero. The larger the number of particles, the smaller the fluctuations and the longer it takes  $f(t)$ to become so small that a fluctuation makes it cross zero and induce a switch of the potential. If $N$ is sufficiently large, this takes in average longer than the optimal periodic switching time. Moreover, since fluctuations are small, this control protocol ends up behaving approximately periodically in time but with switching periods longer than the optimal ones. Hence, it yields a slower motion of the particles than that achieved by the optimal periodic flashing.

The idea underlying the new threshold control policy is to modify that feedback control protocol to prevent the trapping by inducing the switching of the potential when the net force per particle crosses some threshold values that are now different from zero. Thresholds work by removing the long decaying tails in the evolution of the force per particle that produce the trapping and modifying the switching times so that they match the optimal periodic ones for large $N$. This situation is represented in fig.~\ref{evolucion}b. However, to get rid only of these tails while retaining the remaining part, the switching must only occur when the threshold is crossed in the appropriate direction, implying that we need to compute or estimate the sign of the time derivative of the average force per particle.

In the limit $N\to\infty$, evolution of $f(t)$ is well described by $f^\infty_\text{on}(t)$ and $f^\infty_\text{off}(t)$. Hence, to induce a switch at time $\tau_\text{on}$ we should  force the system to change the state of the potential when $f(t)$ attains a threshold value $u_{\text{on}}=f^\infty_\text{on}(\tau_\text{on})$ if the potential is on and $u_{\text{off}}=f^\infty_\text{off}(\tau_\text{off})$ when potential is off.
 However, if the ratchet potential is on for example, we would be interested in switching when the force $f(t)$ crosses $u_{\text{on}}$ {\em and} is decaying towards zero and not when it is increasing, which would be too early. This is readily achieved by  switching off the potential only when $f(t)$ crosses $u_{\text{on}}$ with negative slope. Equivalently, the potential should be switched on only when $f(t)$ crosses $u_{\text{off}}$ with positive slope. In order to translate this into a feedback policy, we need an estimation for this slope in terms of position of the particles. Using Langevin equation \eqref{lang1}, we can write
\begin{equation}
\gamma\dot f(t)= \frac{1}{N}\sum_i  \alpha(t)
F(x_i(t))F\,'(x_i(t)) + \sum_i \xi_i(t)F'(x_i(t)) +
\frac{kT}{2N}\sum_i  F''(x_i(t))
\end{equation}
For a given state of the system $x_i(t)$, a good estimate for the slope of
$f(t)$ is
\begin{equation}
\dot f_{\text{exp}}(x_i(t))\equiv \frac{1}{\gamma N}\sum_i\alpha(t) F(x_i(t))F\,'(x_i(t))+ \frac{kT}{\gamma N}\sum_i F''(x_i(t)) \label{estimate}
\end{equation}
Notice expression \eqref{estimate} can only be used for smooth potentials.

Gathering all the pieces, we can express threshold control policy in the following way
\begin{itemize}

\item \emph{Threshold control}:
\begin{equation} \label{eq.umbral}
\alpha(t) = \begin{cases}
1 & \text{if } f(t)\geq u_{\text{on}}\\
1 & \text{if } u_{\text{off}}<f(t)<u_{\text{on}} \text{ and } \dot f_{\text{exp}}(x_i) \geq0\\
0 & \text{if } u_{\text{off}}<f(t)<u_{\text{on}} \text{ and } \dot f_{\text{exp}}(x_i) <0\\
0 & \text{if } f(t)<u_{\text{off}}
\end{cases}
\end{equation}
where $f(t)$ and $\dot f_{\text{exp}}$ are the net force per particle and the estimate for the time derivative of its average as defined in \eqref{netforce} and \eqref{estimate}, respectively.
\end{itemize}

To compute the thresholds one can simulate Langevin equation \eqref{lang1} for a sufficiently large number of particles and obtain the evolution of the net force $f(t)$ under the optimal periodic switching. Since the number of particles is large, this is almost equivalent to solving Fokker-Planck equation to get $f^{\infty}(t)$. Thresholds are given by $u_{\text{on}}=f^{\infty}_{\text {on}}(\tau_{\text{on}})$ and $u_{\text{off}}=f^{\infty}_{\text{off}}(\tau_{\text{off}})$ and hence, averaging the values of the force at switching times should give a reasonable estimate for either threshold, once the stationary regime has been reached. Doing so for $N=10^7$, gives $u_{\text{on}}\simeq 0.54 kT/L$ and $u_{\text{off}}\simeq -0.30 kT/L$. These are the thresholds used in the simulations whose results are  shown in fig.~\ref{results}.

\section{Results and conclusions}

The response of the system to the three different switching protocols has been measured in terms of the resulting particle current or average center of mass velocity in the stationary regime, as given by
\begin{equation}
\langle \dot x_{\text{cm}}^{\text{st}}\rangle=\lim_{t\to\infty}\frac{\langle x_{\text{cm}}(t)\rangle}{t}
\end{equation}

 Figure \ref{results} shows simulation results for the three strategies, where $\langle \dot x_{\text{cm}}^{\text{st}}\rangle$ is plotted against the number of particles $N$ for $V_0=5kT$. For this potential, the highest center of mass velocity is achieved for switching periods with values $\tau_{\text{on}}=0.06L^2/D$ and $\tau_{\text{off}}=0.05L^2/D$. As already pointed out
instant velocity $\dot x_{\text{cm}}(t)$ maximization yields  higher velocity than the periodic strategy only up to about a thousand particles and the velocity vanishes when $N\rightarrow \infty$.
On the other hand, the feedback control with the aforementioned thresholds  gives a non-vanishing speed for the center of mass that converges to that achieved by the optimal periodic switching, while retaining approximately the same behaviour, for small ensemble sizes, as the maximization of instant velocity. Thus, the feedback strategy described in eq.~\eqref{eq.umbral} performs better than or at least equal to the best open-loop control strategy, no matter the number of particles to be controlled.

It may be surprising at first sight that the feedback control works nicely for small ensemble sizes since it relies on arguments only valid for a large number of particles. However, fluctuations of the force $f(t)$ are so huge for small $N$ that both feedback control policies become effectively the same.

In conclusion, we have devised a feedback control or switching policy that gives higher particle currents than the best periodic strategy for any number of particles in a collective ratchet. The new strategy \eqref{eq.umbral} is a modification of the strategy \eqref{alphacont} already introduced in ref.~\cite{cao}. It is worth noting that the latter is a policy that maximizes the instantaneous value of the current and ends up inducing feeble currents in the stationary regime if the number of particles is sufficiently large. In contrast, the switching on of the potential for the modified strategy produces an instantaneous negative flux and the switching off takes place when particles could still keep moving. However, both sacrifices are eventually compensated and the final result is that particles actually move faster in the long run. Hence, the modified strategy performs better by being less ``greedy''. This same effect has been also pointed out by Cleuren and Van den Broeck in the ratchet inspired paradoxical games \cite{cleu1,cleu2}.

Finally, we would like to remark that, although finding the optimal protocol for $N>1$ is still an unsolved problem. We believe  the threshold control presented here should not perform much worse than the true optimal: it yields approximately the same current for $N=1$ than the true optimal for one particle and on the other hand, it is equivalent to the optimal periodic one for $N\to \infty$.

\acknowledgments

This work is financially supported by Ministerio de Ciencia y Tecnolog\'{i}a (Spain) through grants FIS2004-271 and BFM2003-02547/FISI.

\end{document}